\documentstyle[12pt,epsfig]{article}
\begin{document}
\begin{titlepage}
\begin{flushright}
LNF--00/019(P)\\
UAB--FT--492\\
hep-ph/0008188\\
August 2000
\end{flushright}
\vspace*{1.6cm}

\begin{center}
{\Large\bf Chiral loops and $a_0$(980) exchange in $\phi\rightarrow\pi^0\eta\gamma$}\\
\vspace*{0.8cm}

A.~Bramon$^1$, R.~Escribano$^2$, J.L.~Lucio M.$^{2,3}$, 
M.~Napsuciale$^3$, and G.~Pancheri$^2$\\
\vspace*{0.2cm}

{\footnotesize\it $^1$Departament de F\'{\i}sica,
Universitat Aut\`onoma de Barcelona, 
E-08193 Bellaterra (Barcelona), Spain\\
$^2$INFN-Laboratori Nazionali di Frascati,
P.O.~Box 13, I-00044 Frascati, Italy\\
$^3$Instituto de F\'{\i}sica, Universidad de Guanajuato,
Lomas del Bosque \# 103, Lomas del Campestre, 37150 Le\'on, 
Guanajuato, Mexico}
\end{center}
\vspace*{1.0cm}

\begin{abstract}
The radiative $\phi\rightarrow\pi^0\eta\gamma$ decay is discussed 
emphasizing the effects of the $a_0$(980) scalar resonance which dominates  
the high values of the $\pi^0\eta$ invariant mass spectrum. 
In its lowest part, the proposed amplitude coincides with the reliable and ChPT-inspired
contribution coming from chiral loops. 
The $a_0$(980) resonance is then incorporated exploiting the complementarity between ChPT
and the linear sigma model for this channel. 
The recently reported experimental invariant mass distribution and branching ratio can be
satisfactorily accommodated in our framework. 
For the latter, a value of $B(\phi\rightarrow\pi^0\eta\gamma)$ 
in the range $(0.75$--$0.95)\times 10^{-4}$ is predicted.
\end{abstract}
\end{titlepage}

\section{Introduction}
The 1 GeV energy region is a particularly challenging domain. 
On one side, it is far below from the perturbative QCD regime and, on the other, 
Chiral Perturbation Theory (ChPT) is not expected to make reliable predictions at these 
energy values where resonance effects are known to be present. 
Among the latter, those proceeding by the exchange of the scalar resonances $f_0$(980) 
and $a_0$(980) should dominate their respective channels. 
The controversial nature of these resonances \cite{close:1}
and the poor knowledge of their properties \cite{PDG}
adds then further complexity (and interest) to this 1 GeV energy region. 
Indeed, several proposals have been suggested along the years concerning the 
constitution of these scalars as complex $q q \bar q \bar q$ states  
\cite{jaffe}, $K \bar K$ molecules \cite{weinstein} or ordinary 
$q \bar q$ mesons \cite{niels:1}.

The Novosibirsk CMD-2 and SND Collaborations have reported very recently, among others, 
the branching ratio and the $\pi^0\eta$ invariant mass distribution for the 
$\phi\rightarrow\pi^0\eta\gamma$ decay. 
For the branching ratio, the CMD-2 Collaboration reports 
$B(\phi\rightarrow\pi^0\eta\gamma)=(0.90\pm 0.24\pm 0.10)\times 10^{-4}$ \cite{CMD-2:1}, 
while the SND result is, consistently,  
$B(\phi\rightarrow\pi^0\eta\gamma)=(0.88\pm 0.14\pm 0.09)\times 10^{-4}$ \cite{SND:1}.
The observed invariant mass distribution shows a significant enhancement at large 
$\pi^0\eta$ invariant mass that, according to Refs.~\cite{CMD-2:1,SND:1}, 
could be interpreted as a manifestation of a sizeable contribution of the 
$a_0(980)\gamma$ intermediate state.
This and other radiative $\phi$ decays are also expected to be intensively investigated 
at the Frascati $\phi$-factory DA$\Phi$NE \cite{daphne95:juliet}.

On the theoretical side, the $V \rightarrow P^0 P^0\gamma$ decays have been considered
by a number of authors \cite{close:1,achasov:1,bramon:1,lucio:1,oset:1}.
In particular, it has been shown that the intermediate vector meson contributions to 
$\phi\rightarrow\pi^0\eta\gamma$ lead to a small
$B(\phi\rightarrow\pi^0\eta\gamma)_{\rm VMD}=5.4 \times 10^{-6}$ \cite{BGP:1}, 
whereas a chiral loop model closely linked to standard ChPT predicts 
$B(\phi\rightarrow\pi^0\eta\gamma)_\chi=3.0 \times 10^{-5}$ \cite{BGP:2}. 
Needless to say, the scalar resonance effects and, in particular, the resonance pole 
associated to the $a_0$(980) were not contemplated in these two approaches. 
The recent experimental data from Novosibirsk 
---for both the branching ratio and the $\pi^0\eta$ invariant mass spectrum showing 
an enhancement around the $a_0$(980) mass---  
seem therefore to disfavour these predictions based on vector meson exchange and/or a
simple extrapolation of ChPT ideas.

If we rely on the resonance picture, it is clear that the $a_0(980)$ scalar meson 
---lying just below the $\phi$ mass and having the appropriate quantum numbers--- 
should play an important r\^ole in the $\phi\rightarrow\pi^0\eta\gamma$ decay. 
Several theoretical attempts to describe the effects of scalars in $\phi$ radiative 
decays have appeared so far. Among others, we would like to refer to the 
``no structure" model \cite{bramon:2}, to the $K^+K^-$ model \cite{close:1,lucio:1}, 
where the $\phi\rightarrow a_0\gamma$ amplitude is generated through a loop of charged 
kaons, and to the chiral unitary approach $(U\chi PT)$ \cite{oset:1}, where the decay 
$\phi\rightarrow\pi^0\eta\gamma$ occurs through a loop of charged kaons that subsequently 
annihilate into $\pi^0\eta\gamma$.
In the two former cases the scalar resonances are included {\it ad hoc}  
while in the latter they are generated dynamically by unitarizing the one-loop amplitudes.

In this letter, we are mainly interested in incorporating scalar resonances and their 
pole effects into a ChPT inspired context \cite{ecker:1}. 
While vector and axial-vector resonances can be included in a transparent and successful way, 
offering some theoretical basis to conventional vector meson dominance (VMD) ideas 
\cite{ecker:2}, the incorporation of scalar resonances has been more ambiguous and less
successful up to now \cite{ecker:1}.
In order to take explicitly into account scalar resonances and their pole effects, 
we propose to use the linear sigma model (L$\sigma$M). 
This will allow us to take advantage of the common origin of ChPT and the L$\sigma$M to
improve the  chiral loop predictions for $V\rightarrow P^0 P^0\gamma$ exploiting the
complementarity  of both approaches for these specific processes.
On one side, ChPT is the established theory of the pseudoscalar interactions at low energy. 
However, it is not reliable at energies of a typical vector meson mass and, 
as just stated, scalar resonance poles are not explicitly included. 
As a consequence, ChPT inspired loop models can give rough estimates for 
$B(V\rightarrow P^0 P^0\gamma)$ but will hardly be able to reproduce the observed 
enhancements in the invariant mass spectra. 
On the other side, the L$\sigma$M is a much simpler model dealing similarly with 
pseudoscalar interactions but incorporating {\it scalar} resonances in a systematic and 
definite way. 
Thanks to this, the L$\sigma$M should be able to reproduce the resonance peaks in the 
spectra and, although it does not provide a systematical framework for the pseudoscalar
meson physics, this model could be of relevance in describing the scalar resonances when
linked to a well established ChPT context. 
In order to show in detail the proposed framework, we will focus our attention on the 
$\phi\rightarrow\pi^0\eta\gamma$ decay mode. 
Other decay modes are somewhat more involved and will be analyzed in forthcoming work.

\section{$\phi\rightarrow\pi^0\eta\gamma$ and chiral loops}
\label{sectChPT}

The vector meson initiated $V\rightarrow P^0 P^0\gamma$ decays cannot be treated in strict 
Chiral Perturbation Theory (ChPT). This theory has to be extended to incorporate on-shell 
vector meson fields. At lowest order, this may be easily achieved by means of the 
${\cal O}(p^2)$ ChPT Lagrangian:
\begin{equation}
\label{ChPTlag}
{\cal L}_2=
\frac{f^2}{4}\langle D_\mu U^\dagger D^\mu U+U^\dagger\chi +\chi^\dagger U\rangle\ ,
\end{equation}
where $f=f_\pi=92.4$ MeV at this order, $U=\exp(i\sqrt{2}P/f)$ with $P$ being the usual
pseudoscalar nonet matrix, and $\chi=2 B_0 {\cal M}$ with 
${\cal M}=\mbox{diag}(m_u, m_d, m_s)$.
The covariant derivative, now enlarged to include vector mesons, is defined as 
$D_\mu U=\partial_\mu U -i e A_\mu [Q,U] - i g [V_\mu,U]$, with 
$Q=\mbox{diag}(2/3, -1/3, -1/3)$ being the quark charge matrix and $V_\mu$ the additional 
matrix containing the nonet of ideally mixed vector meson fields. The diagonal elements of 
$V$ are $(\rho^0+\omega)/\sqrt{2}, (-\rho^0+\omega)/\sqrt{2}$ and $\phi$, thus following 
the same conventional normalization as for the pseudoscalar nonet matrix $P$.

There is no tree-level contribution from this Lagrangian to the  
$\phi\rightarrow\pi^0\eta\gamma$ amplitude and at the one-loop level one needs to
compute the set of diagrams shown in Ref.~\cite{BGP:2}.
A straightforward calculation leads to the following {\it finite} amplitude for 
$\phi(q^\ast,\epsilon^\ast)\rightarrow \pi^0(p)\eta(p^\prime)\gamma(q,\epsilon)$
(see Ref.~\cite{BGP:2} for further details):
\begin{equation}
\label{AphiChPT}
\begin{array}{rl}
{\cal A}(\phi\rightarrow \pi^0\eta\gamma)_\chi &=\ 
\frac{eg}{2\pi^2 m^2_{K^+}}
(\epsilon^\ast\epsilon\ q^\ast q - \epsilon^\ast q\ \epsilon q^\ast)
L(m^2_{\pi^0\eta})\\[2ex]
&\times\ {\cal A}(K^+K^-\rightarrow\pi^0\eta)_\chi\ ,
\end{array}
\end{equation}
where $\epsilon^\ast\epsilon\ q^\ast q - \epsilon^\ast q\ \epsilon q^\ast$
makes the amplitude Lorentz- and gauge-invariant, 
$m^2_{\pi^0\eta}\equiv s\equiv (p+p^\prime)^2=(q^\ast -q)^2$ is the invariant mass of the 
final pseudoscalar system and $L(m^2_{\pi^0\eta})$ is the loop integral function defined as
\begin{equation}
\label{L}
\begin{array}{rl}
L(m^2_{\pi^0\eta}) &=\ 
\frac{1}{2(a-b)}-
\frac{2}{(a-b)^2}\left[f\left(\frac{1}{b}\right)-f\left(\frac{1}{a}\right)\right]\\[2ex]
&+\ \frac{a}{(a-b)^2}\left[g\left(\frac{1}{b}\right)-g\left(\frac{1}{a}\right)\right]\ ,
\end{array}
\end{equation}
where
\begin{equation}
\label{f&g}
\begin{array}{l}
f(z)=\left\{
\begin{array}{ll}
-\left[\arcsin\left(\frac{1}{2\sqrt{z}}\right)\right]^2 & z>\frac{1}{4}\\[1ex]
\frac{1}{4}\left(\log\frac{\eta_+}{\eta_-}-i\pi\right)^2 & z<\frac{1}{4}
\end{array}
\right.\\[5ex]
g(z)=\left\{
\begin{array}{ll}
\sqrt{4z-1}\arcsin\left(\frac{1}{2\sqrt{z}}\right) & z>\frac{1}{4}\\[1ex]
\frac{1}{2}\sqrt{1-4z}\left(\log\frac{\eta_+}{\eta_-}-i\pi\right) & z<\frac{1}{4}
\end{array}
\right.
\end{array}
\end{equation}
and
$\eta_\pm=\frac{1}{2}(1\pm\sqrt{1-4z})$, $a=\frac{m^2_\phi}{m^2_{K^+}}$ and 
$b=\frac{m^2_{\pi^0\eta}}{m^2_{K^+}}$.
The coupling constant $g$ comes from the strong amplitude
${\cal A}(\phi\rightarrow K^+K^-)=g\epsilon^\ast (p_+-p_-)$ with $|g|=4.59$
to agree with $\Gamma (\phi\rightarrow K^+K^-)_{\rm exp}= 2.19$ MeV.
The latter is the part beyond standard ChPT which we have fixed phenomenologically.
The four-pseudoscalar amplitude is instead a standard ChPT amplitude\footnote{
${\cal A}(K^+K^-\rightarrow\pi^0\eta_8)_\chi=
\frac{\sqrt{3}}{4 f_\pi^2}\left(m^2_{\pi^0\eta}-\frac{4}{3}m^2_K\right)$ 
if only the $\eta_8$ contribution is taken into account as in Ref.~\protect\cite{BGP:2}.}
which is found to depend linearly on the variable $s=m^2_{\pi^0\eta}$:
\begin{equation}
\label{A4PChPTphys}
{\cal A}(K^+K^-\rightarrow\pi^0\eta)_\chi=\frac{1}{\sqrt{6}f_\pi^2}
\left(m^2_{\pi^0\eta}-\frac{10}{9}m^2_K+\frac{1}{9}m^2_\pi\right)\ .
\end{equation}
In the calculation of the decay amplitudes (\ref{AphiChPT}) and (\ref{A4PChPTphys})  
we have introduced $\eta$-$\eta^\prime$ mixing effects. 
As it is well known, a rigorous and general extension of $SU(3)$ ChPT to include the ninth 
pseudoscalar meson $\eta_0$ is not straightforward and requires the introduction of new 
terms in the chiral Lagrangian \cite{UB}.
However, if one relies on classical arguments based on nonet symmetry, a phenomenologically 
successful description of the $\eta$-$\eta^\prime$ system is achieved \cite{bijnens:1}.
The $\eta$-$\eta^\prime$ mixing angle is then found to be compatible with 
$\theta_P=\arcsin(-1/3)\simeq -19.5^\circ$, quite in agreement with recent phenomenological 
estimates \cite{thetaP}. 
In Sect.~\ref{sectLsigmaM}, it will be shown that this choice for the 
$\eta$-$\eta^\prime$ mixing angle greatly simplifies the calculation of the 
$\phi\rightarrow\pi^0\eta\gamma$ amplitude in the L$\sigma$M, 
and reduces up to a minimum the number of free-parameters. 

\begin{figure}[t]
%\centerline{\epsfig{file=ChPTspectrum.eps,height=4cm}}
\centerline{\epsffile{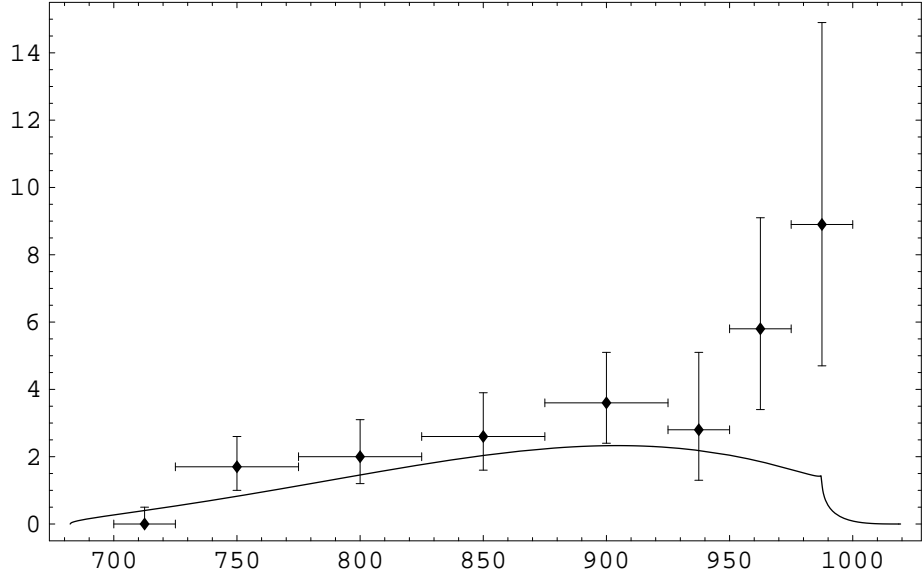}}
\caption{$dB(\phi\rightarrow\pi^0\eta\gamma)/dm_{\pi^0\eta}\times 10^7\ \mbox{MeV}^{-1}$
as a function of the $m_{\pi^0\eta}$ invariant mass in a chiral loop model. 
Experimental data are taken from Ref.~\protect\cite{SND:1}.}
\label{figChPT}
\end{figure}

The invariant mass distribution for the $\phi\rightarrow\pi^0\eta\gamma$ decay 
is predicted to be given by the following spectrum (see Fig.~\ref{figChPT}):
\begin{equation}
\label{dGChPT}
\begin{array}{rl}
\frac{d\Gamma(\phi\rightarrow\pi^0\eta\gamma)_\chi}{dm_{\pi^0\eta}}&=\ 
\frac{\alpha}{192\pi^5}\frac{g^2}{4\pi}\frac{m^4_\phi}{m^4_{K^+}}
\frac{m_{\pi^0\eta}}{m_\phi}\left(1-\frac{m^2_{\pi^0\eta}}{m^2_\phi}\right)^3
\sqrt{1-2\frac{m^2_{\pi^0}+m^2_\eta}{m^2_{\pi^0\eta}}+
\left(\frac{m^2_\eta-m^2_{\pi^0}}{m^2_{\pi^0\eta}}\right)^2}\\[2ex]
&\times\ |L(m^2_{\pi^0\eta})|^2
|{\cal A}(K^+K^-\rightarrow\pi^0\eta)_\chi|^2\ .
\end{array}
\end{equation}
Integrating Eq.~(\ref{dGChPT}) over the whole physical region one obtains for the 
branching ratio:
\begin{equation}
\label{BRChPT}
B(\phi\rightarrow\pi^0\eta\gamma)_\chi=0.47\times 10^{-4}\ .
\end{equation}
As expected, Fig.~\ref{figChPT} shows that our chiral loop approach gives a reasonable 
prediction for the lower part of the spectrum but fails to reproduce the observed 
enhancement in its higher part, where $a_0(980)$-resonance effects 
(ignored up to this point of our approach) should manifest. 
As a consequence, the predicted branching ratio turns out to be below the experimental 
value by about a factor of 2.

\section{Improved approach to $\phi\rightarrow\pi^0\eta\gamma$}
\label{sectLsigmaM}

To analyze the scalar resonance effects in the $V\rightarrow P^0 P^0\gamma$ decay 
amplitudes, the linear sigma model (L$\sigma$M) \cite{oldsigmamodel} 
will be shown to be particularly appropriate.
It is a well-defined $U(3)\times U(3)$ chiral model which incorporates {\it ab initio} both 
the nonet of pseudoscalar mesons together with its chiral partner, the scalar mesons nonet. 
Recently, the model has  been resurrected as a framework to study the implications of chiral 
symmetry for the controversial scalar sector of QCD, and some variations of the basic 
L$\sigma$M Lagrangian have shown to be phenomenologically rather 
successful \cite{mauro,niels,evgenii}.

In this context, the $V\rightarrow P^0 P^0\gamma$ decays 
proceed through a loop of charged pseudoscalar mesons emitted by the initial vector. 
Due to the additional emission of a photon, these charged pseudoscalar pairs with the initial 
$J^{PC} = 1^{--}$ quantum numbers can rescatter into $J^{PC} = 0^{++}$ pairs of charged or 
neutral pseudoscalars.  
The scalar resonances are expected to play an essential r\^ole in this rescattering process 
and the L$\sigma$M seems mostly appropriate to fix the corresponding amplitudes.

\begin{figure}[t]
\centerline{\epsfig{file=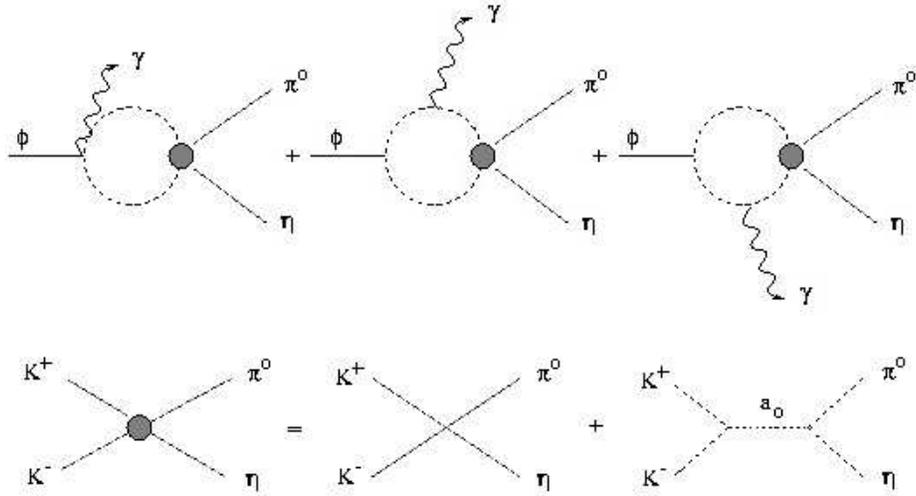,height=8cm}}
%\centerline{\epsffile{LsigmaMloops.eps}}
\caption{One-loop Feynman diagrams for $\phi\rightarrow\pi^0\eta\gamma$ in the L$\sigma$M 
for a $\eta$-$\eta^\prime$ mixing angle $\theta_P=\arcsin(-1/3)\simeq -19.5^\circ$.}
\label{figloops}
\end{figure}

Several simplifications happen when one considers the $\phi\rightarrow\pi^0\eta\gamma$ 
decay mode. 
As in the analysis of Sect.~\ref{sectChPT}, contributions from charged pions in the loops 
are highly suppressed because they involve the isospin violating and OZI--rule forbidden 
$\phi\pi\pi$ coupling; hence, the dominant contributions arise exclusively from loops of 
charged kaons. 
The subsequent rescattering of these charged kaon pairs into the final $J^{PC} = 0^{++}$ 
$\pi^0\eta$ state is then quite simple. 
Indeed, the L$\sigma$M amplitude for $K^+K^-\rightarrow\pi^0\eta$ contains a contact term, 
a term with an $a_0$ exchanged in the $s$-channel and two terms with a $\kappa$ 
({\it i.e.~}the strange $I=1/2$ scalar resonance) exchanged in the $t$- and $u$-channels. 
However, the latter $\kappa$-exchange contributions are absent for an $\eta$-$\eta^\prime$ 
mixing angle $\theta_P=\arcsin(-1/3)\simeq -19.5^\circ$ since the $g_{\kappa K\eta}$ 
coupling constant appearing in one of the $\kappa$ vertices vanishes. 
The calculation is then reduced to the diagrams shown in Fig.~\ref{figloops}, 
and it is thus much simpler for this particular (but phenomenologically 
acceptable \cite{thetaP}) choice of the $\eta$-$\eta^\prime$ mixing angle. 
Moderate departures from this value translate into weak $g_{\kappa K\eta}$ 
couplings\footnote{The coupling constant 
$g_{\kappa K\eta}$ is proportional to $\sin\left[\theta_P-\arcsin(-1/3)\right]$ and its 
dependence on the $\eta$-$\eta^\prime$ mixing angle for values around 
$\theta_P=\arcsin(-1/3)\simeq -19.5^\circ$ is soft.} 
appearing in the $\phi\rightarrow\pi^0\eta\gamma$ amplitude. 
Their effects seem to be small (see below) for the present and expected levels of 
experimental accuracy and for this kind of processes governed by poorly known scalar 
resonances. 
More importantly, the absence of $\kappa$ contributions makes our predictions more  
definite and solid since we avoid one of the major uncertainties affecting the scalar 
nonet dynamics, namely, the mass of its strange members. Indeed, recent analysis by 
various authors require a light 
$\kappa$(900) \cite{jaffe,mauro,schechter,ishida,mike}, 
while other authors deny the existence of a low mass pole  \cite{cherry} and identify the 
PDG $K_0^\ast(1430)$ state with the strange member of the lowest lying scalar nonet  
\cite{niels,evgenii}. 

A straightforward calculation of the $\phi\rightarrow\pi^0\eta\gamma$ decay amplitude 
leads to an expression identical to that in Eq.~(\ref{AphiChPT}) but with the 
four-pseudoscalar amplitude now computed in a L$\sigma$M context. 
In this case, the amplitude is just
\begin{equation}
\label{A4PLsigmaM}
{\cal A}(K^+K^-\rightarrow\pi^0\eta)_{\mbox{\scriptsize L$\sigma$M}}=
g_{K^+K^-\pi^0\eta}-\frac{g_{a_0K^+K^-}g_{a_0\pi^0\eta}}
{D_{a_0}(m^2_{\pi^0\eta})}\ ,
\end{equation}
where 
$D_{a_0}(m^2_{\pi^0\eta})$ is the $a_0$ propagator and the coupling constants 
are\footnote{See Ref.~\protect\cite{mauro} for a detailed calculation of these coupling 
constants in the L$\sigma$M.}
\begin{equation}
\label{gLsigmaM}
\begin{array}{l}
g_{K^+K^-\pi^0\eta}=\frac{g_{a_0\pi^0\eta}}{2f_K}=
-\sqrt{\frac{2}{3}}\frac{m^2_{a_0}-m^2_\eta}{2f_Kf_\pi}\ ,\\[2ex]
g_{a_0K^+K^-}=-\frac{m^2_{a_0}-m^2_K}{2f_K}\ .
\end{array}
\end{equation}
This amplitude can then be rewritten as
\begin{equation}
\label{A4PLsigmaMphys}
{\cal A}(K^+K^-\rightarrow\pi^0\eta)_{\mbox{\scriptsize L$\sigma$M}}= 
\frac{1}{\sqrt{6}f_Kf_\pi}(m^2_{\pi^0\eta}-m^2_K)\times
\frac{m^2_\eta-m^2_{a_0}}{D_{a_0}(m^2_{\pi^0\eta})}\ .
\end{equation}

We would like to make a few remarks on the four-pseudoscalar amplitude 
in Eq.~(\ref{A4PLsigmaMphys}) and compare it with the ChPT amplitude 
in Eq.~(\ref{A4PChPTphys}):
\begin{itemize}
\item[i)] For $m_{a_0}\rightarrow\infty$ and ignoring $SU(3)$-breaking in the pseudoscalar 
masses and decay constants, the L$\sigma$M amplitude (\ref{A4PLsigmaMphys}) reduces to the 
ChPT one (\ref{A4PChPTphys}). 
The former consists of a constant four-pseudoscalar vertex plus a second term whose 
$s$-dependence is generated by the $s\equiv m^2_{\pi^0\eta}$ piece in the $a_0$ propagator 
$D_{a_0}(s = m^2_{\pi^0\eta})$, as shown in Eq.~(\ref{A4PLsigmaM}). 
Their sum (see Eq.~(\ref{A4PLsigmaMphys})) in the good $SU(3)$ and $m_{a_0}\rightarrow\infty$ 
limits ends up with an amplitude which is linear in $s=m^2_{\pi^0\eta}$ and mimics perfectly 
the effects of the derivative and massive terms in the ChPT Lagrangian (\ref{ChPTlag}) 
leading  respectively to the two terms in the ChPT amplitude (\ref{A4PChPTphys}). 
This, we believe, is the main virtue of our approach and makes the use of the L$\sigma$M
reliable at least for amplitudes like ours where $s$-channel exchange plays the 
main r\^ole and $(t,u)$-channel exchange can be ignored. 
\item[ii)] The L$\sigma$M and ChPT yield slightly different amplitudes 
in the $m_{a_0}\rightarrow\infty$ limit because of the way $SU(3)$-symmetry is broken 
in the two approaches.
In the case of the L$\sigma$M  \cite{mauro,niels,evgenii}, a non $SU(3)$ symmetric choice of
the vacuum expectation values makes simultaneously $m_\pi^2\neq m_K^2$ and $f_\pi\neq f_K$, 
whereas in ChPT $m_\pi^2\neq m_K^2$ is already present in the lowest order Lagrangian while 
$f_\pi\neq f_K$ is only achieved at higher orders.
%Thus, most of these SU(3)-breaking effects are beyond control in our present approach; 
\item[iii)] In the $\phi\rightarrow\pi^0\eta\gamma$ decay, the threshold for $\pi^0\eta$ 
production is not far from the mass of the $a_0$(980) and that makes crucial 
the incorporation of the $a_0$ in an explicit way.
Due to the presence of the full propagator $D_{a_0}(s)$, as in Eq.~(\ref{A4PLsigmaMphys}),
such an amplitude ---closely linked to that from ChPT and thus expected to be able to 
account for the lowest part of the $\pi^0\eta$ mass spectrum--- should also be able to 
reproduce the effects of the $a_0$ pole at higher $\pi^0\eta$ invariant mass values. 
\item[iv)] The need for the $a_0$ propagator introduces, however, some uncertainties in our 
treatment. Indeed, the opening of the $K\bar{K}$ channel near the $a_0$(980) mass has 
motivated the use of different expressions for $D_{a_0}(s)$. 
A first possibility consists in using a Breit-Wigner propagator with an energy dependent 
width (to incorporate the known kinematic corrections):
\begin{equation}
\label{BW}
D_{a_0}(s) = s-m^2_{a_0}+i\sqrt{s}\,\Gamma_{a_0}(s)\ ,
\end{equation}
where
\begin{equation}
\label{Gamma}
\begin{array}{rl}
\Gamma_{a_0}(s) &=\ \frac{g^2_{a_0\pi^0\eta}}{16\pi\sqrt{s}} 
\sqrt{\left[1-\frac{(m_{\pi^0} + m_\eta)^2}{s}\right]
      \left[1-\frac{(m_{\pi^0} - m_\eta)^2}{s}\right]}\, 
\theta (\sqrt{s}-(m_{\pi^0} + m_\eta))\\[2ex]
&+\ \frac{g^2_{a_0 K^+ K^-}}{16\pi\sqrt{s}}\sqrt{1-\frac{4m_{K^+}^2}{s}}\,
\theta (\sqrt{s}-2m_{K^+})\\[2ex]
&+\ \frac{g^2_{a_0 K^0 \bar K^0}}{16\pi\sqrt{s}}\sqrt{1-\frac{4m_{K^0}^2}{s}}\,
\theta (\sqrt{s}-2m_{K^0})\ .
\end{array}
\end{equation}
Another interesting and widely accepted option was proposed by Flatt\'e time ago 
specifically for the two-channel $a_0$ resonance \cite{flatte}. 
The relative narrowness of the observed ${\pi\eta}$ peak around 980 MeV is then explained by 
the action of unitarity and analyticity at the $K\bar{K}$ threshold. This amounts to 
extend the preceding formulae below the $K \bar K$ threshold to include the now 
purely imaginary kaon contributions. 
\end{itemize}
Due to these distinct possibilities to deal with the $a_0$ propagator, 
as well as to other differences introduced when implementing and fitting the basic L$\sigma$M 
Lagrangian by several authors, a set of predictions can be  obtained for the 
four-pseudoscalar amplitude (\ref{A4PLsigmaMphys}). 
In turn, these various amplitudes have to substitute the four-pseudoscalar ChPT amplitude in 
Eq.~(\ref{dGChPT}) to finally obtain the corresponding invariant mass distributions of the 
$\phi\rightarrow\pi^0\eta\gamma$ decay mode\footnote{Since all the four-pseudoscalar
amplitudes we are considering depend only on the variable $s$, they factorize out of the 
loop integration and the structure of Eq.~(\ref{dGChPT}) is fully preserved.}.  
Our purpose hereafter is to briefly discuss a few of these treatments in order to show that 
the observed properties for this specific decay can be accommodated in our ChPT- and 
L$\sigma$M-inspired approach.

\begin{figure}[t]
%\centerline{\epsfig{file=LsigmaMspectrum.eps,height=4cm}}
\centerline{\epsffile{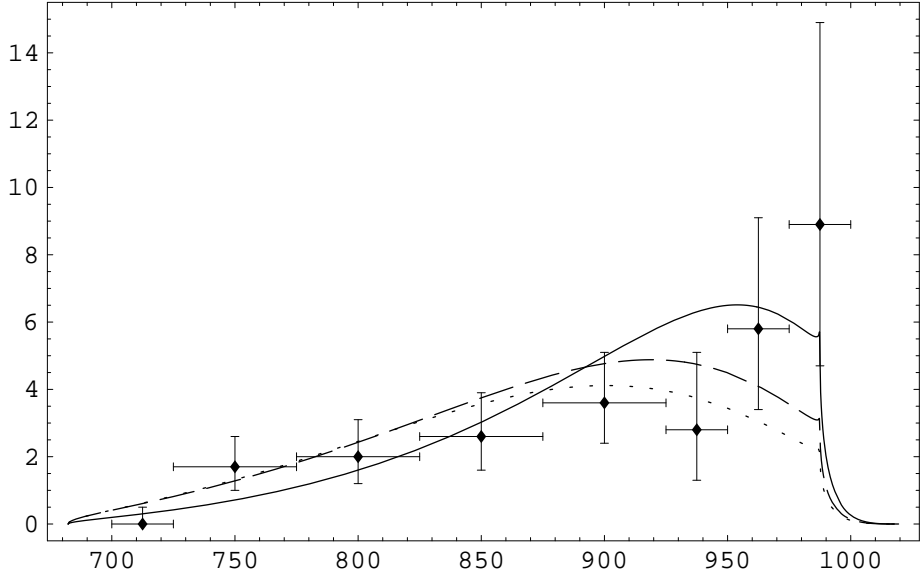}}
\caption{$dB(\phi\rightarrow\pi^0\eta\gamma)/dm_{\pi^0\eta}\times 10^7\ \mbox{MeV}^{-1}$
as a function of the $m_{\pi^0\eta}$ invariant mass in the L$\sigma$M.
The dotted, dashed and solid lines correspond to the versions of the L$\sigma$M proposed 
by Refs.~\protect\cite{mauro,niels,evgenii} respectively. 
Experimental data are taken from Ref.~\protect\cite{SND:1}.}
\label{figLsigmaM}
\end{figure}

We start our discussion along the lines of Ref.~\cite{mauro} taking for the $a_0$ propagator
the simple Breit-Wigner prescription in Eq.~(\ref{BW}). The use of this propagator for the
L$\sigma$M amplitude Eq.~(\ref{A4PLsigmaMphys}) and its insertion in Eq.~(\ref{dGChPT})
predicts the $m_{\pi^0\eta}$ invariant mass spectrum shown by the dotted line in 
Fig.~\ref{figLsigmaM}. 
Integrating over the whole physical region leads to the branching ratio
\begin{equation}
\label{BRmauro}
B(\phi\rightarrow\pi^0\eta\gamma)_{\mbox{\scriptsize L$\sigma$M\cite{mauro}}}=
0.80\times 10^{-4}\ ,
\end{equation}
in agreement with the experimental data \cite{CMD-2:1,SND:1}.
However, since the simple expression used for the $a_0$ propagator implies a large 
$a_0$-width ($\Gamma_{a_0\rightarrow\pi\eta}\simeq 460$ MeV \cite{mauro}), 
the desired enhancement in the invariant mass spectrum appears in its central part rather 
than around the $a_0$ peak.   

This unpleasant feature is easily corrected when turning to the proposal 
by T\"ornqvist \cite{niels}. 
Indeed, a Gaussian form factor related to the finite size of physical mesons and depending on 
the final CM-momentum is introduced to describe the decays of scalar resonances in this 
approach. As a result, the decay width of $a_0$(980) into $\pi^0\eta$ is reduced 
($\Gamma_{a_0\rightarrow\pi\eta}\simeq 273$ MeV \cite{niels}) without 
affecting that of $a_0$(980) into $K\bar{K}$. This fact produces an enhancement 
in the spectrum for the higher values of the 
$m_{\pi^0\eta}$ invariant mass, as shown by the dashed line  
in Fig.~\ref{figLsigmaM}. 
The integrated branching ratio is then predicted to be 
\begin{equation}
\label{BRniels}
B(\phi\rightarrow\pi^0\eta\gamma)_{\mbox{\scriptsize L$\sigma$M\cite{niels}}}=
0.90\times 10^{-4}\ ,
\end{equation}
in good agreement with the experimental data \cite{CMD-2:1,SND:1}. 
A possible difficulty of this approach at the phenomenological level is that it predicts
an $\eta$-$\eta^\prime$ mixing angle of $\theta_P \simeq -5.0^\circ$ \cite{niels}, 
considerably less negative than the usually accepted value 
($\theta_P$ between $-20^\circ$ and $-15^\circ$ \cite{thetaP}) and the value required in
our simplified analysis $\theta_P\simeq -19.5^\circ$.
This allows for an estimate of the typical errors introduced when neglecting 
$\kappa$-exchange in the $t$- and $u$-channels as compared to $a_0$-exchange in $s$-channel. 
The significant factor is the ratio of coupling constants 
$r\equiv (g_{\kappa^{\pm} K^{\pm}\eta}\,g_{\kappa^{\pm} K^{\pm}\pi^0})/ 
(g_{a_0\pi^0\eta}\,g_{a_0 K^+ K^-})$, which vanishes in the ideal situation where 
$\theta_P \simeq -19.5^\circ$ but takes the value $r\simeq -1/3$ for 
$\theta_P \simeq -5.0^\circ$ if one ignores $SU(3)$-breaking corrections. 
To the smallness of $r\simeq -1/3$ one has to add the fact that the amplitude for the almost
on-shell $a_0$-exchange in the $s$-channel is mainly imaginary and does not interfere with 
the almost real amplitude for off-shell $\kappa$-exchange in the $t$- and $u$-channels.
As a result, the error in Eq.~(\ref{BRniels}) introduced by neglecting this latter 
contribution can be estimated to be below some 10\% even for such unusual values of 
$\theta_P\simeq -5^\circ$. 

None of these drawbacks are encountered when turning to the treatment proposed by 
Shabalin \cite{evgenii}.
In the fitting procedure adopted by this author no attempt is made to fix the 
$\eta$-$\eta^\prime$ mixing angle within the model.
Thanks to this, one minimizes the uncertainties associated with the incorporation of the
ninth pseudoscalar meson $\eta_0$ via the axial anomaly term.
The value of the $\eta$-$\eta^\prime$ mixing angle is then fixed outside the model to its
phenomenologically preferred value $\theta_P\simeq -19.5^\circ$.
Another relevant feature of Shabalin's approach is the introduction of the well-known
Flatt\'e corrections to the $a_0(980)$ propagator.
The $a_0$-width is then drastically reduced from the uncorrected value 
$\Gamma_{a_0}\simeq 304$ MeV to a more acceptable visible width of 
$\Gamma_{a_0}\simeq 65$ MeV.
With all this information taken from Ref.~\cite{evgenii}, our approach predicts
the $m_{\pi^0\eta}$ invariant mass spectrum shown by the solid line in Fig.~\ref{figLsigmaM}
and the integrated branching ratio 
\begin{equation}
\label{BRevgenii}
B(\phi\rightarrow\pi^0\eta\gamma)_{\mbox{\scriptsize L$\sigma$M\cite{evgenii}}}=
0.93\times 10^{-4}\ .
\end{equation}
Both the spectrum and the branching ratio are in nice agreement with the 
experimental data \cite{CMD-2:1,SND:1}.
The fact that Shabalin's model incorporates the Flatt\'e corrections to the 
$a_0$ resonant shape \cite{flatte} has played a substantial r\^ole in this achievement.

\section{Conclusions}
The main aim of the present letter has been to propose and discuss an amplitude for 
the radiative $\phi\rightarrow\pi^0\eta\gamma$ decay exploiting the complementarity 
between ChPT and L$\sigma$M ideas. 
Thanks to the latter, our amplitude contains the full propagator of the $a_0$(980) scalar 
resonance which dominates the higher part of the $\pi^0\eta$ invariant mass spectrum. 
In the low invariant mass region, where ChPT is expected to work quite reliably, the 
proposed amplitude is shown to coincide with that coming from a chiral loop calculation. 
This, we believe, makes reliable our approach to the $V\rightarrow P^0P^0\gamma$ dynamics and,
in particular, to the $\phi\rightarrow\pi^0\eta\gamma$ decay mode for which some
simplifying conditions hold and lead to a simple and well-defined amplitude.
Then our predictions depend only on a reduced number of parameters which, in 
principle, can be extracted from independent data. 
Some of these data refer to scalar meson properties which are not well established and thus 
affect the accuracy of our predictions although by no means in a drastic way.     

We can safely conclude that all the reported properties for the 
$\phi\rightarrow\pi^0\eta\gamma$ decay mode can be accommodated in our approach. 
The branching ratio is predicted to be in the range 
$BR(\phi\rightarrow\pi^0\eta\gamma)=(0.75$--$0.95)\times 10^{-4}$, compatible with the 
present available data.
Similarly, the measured $\pi^0\eta$ invariant mass spectrum is reproduced by our amplitude 
in a reasonable way. 
The uncertainties affecting these predictions suggest that further tests and more refined 
analyses are needed, particularly when the higher accuracy data from ongoing experiments 
will be available. 
This should contribute to clarify one of the most controversial aspects of hadron physics: 
the scalar states around 1 GeV.

\section*{Acknowledgements}
We would like to thank A.~Farilla 
for helpful comments and clarifying discussions.
Work partly supported by the EEC, TMR-CT98-0169, EURODAPHNE network.
Work also partly supported by the CONACyT (project I27604-E).
J.L.~Lucio M.~acknowledges partial financial support from CONACyT and
CONCyTEG.

\end{document}